\documentclass[12pt,showpacs,showkeys,amsmath,aps,prc]{revtex4}

\usepackage{amsfonts} 
\usepackage{mathrsfs}
\usepackage{subfigure}
\usepackage{supertabular}
\usepackage{color,graphicx}
\usepackage[bookmarksopen]{hyperref}

\def  \p    {\pi}

\def  \ra   {\rightarrow}

\def  \veps {\varepsilon}

\def  \del  {\partial}

\def  \bef  {\begin{figure}}
\def  \eef  {\end{figure}}
\def  \be   {\begin{equation}}
\def  \ee   {\end{equation}}
\def  \ba   {\begin{array}}
\def  \ea   {\end{array}}
\def  \bea  {\begin{eqnarray}}
\def  \eea  {\end{eqnarray}}
\def  \beq  {\begin{eqnarray}}
\def  \eeq  {\end{eqnarray}}
\def  \nn   {\nonumber}
\def  \bd   {\begin{displaymath}}
\def  \ed   {\end{displaymath}}
\def  \bse  {\begin{subequations}}
\def  \ese  {\end{subequations}}
\def  \bwt  {\begin{widetext}}
\def  \ewt  {\end{widetext}}

\def  \ba   {{\bf{a_1}}}

\begin{document}
\title{Spin susceptibility of degenerate quark matter}
\author {Kausik Pal}
\email {kausik.pal@saha.ac.in}
\author {Abhee K. Dutt-Mazumder}
\affiliation {High Energy Physics Division, Saha Institute of Nuclear Physics,
 1/AF Bidhannagar, Kolkata 700064, India.}

\medskip

\begin{abstract}

The expression for the spin susceptibility $\chi$ of degenerate quark matter 
is derived with corrections upto $ {\cal O}(g^4\ln g^2)$. It is shown 
that at low density, $\chi^{-1}$ changes sign and turns negative indicating a ferromagnetic phase transition.
To this order, we also calculate sound velocity $c_1$ and incompressibility $K$
with arbitrary spin polarization. The estimated values of $c_1$ and $K$ 
show that the equation of state of the polarized matter is stiffer than the unpolarized one. Finally we determine the finite temperature 
corrections to the exchange energy and derive corresponding results for 
the spin susceptibility.
\end{abstract}
\vspace{0.08 cm}



\maketitle

\section{Introduction}

One of the active areas of high energy physics research has been
exploration of the so called Quantum Chromodynamics (QCD) phase
diagram. In particular, with the advent of ultrarelativistic heavy ion 
beams at RHIC and CERN and with the upcoming facilities of GSI where
compressed baryonic matter is expected to be produced, such studies
have assumed special importance. Beside the laboratory experiments,
various astrophysical objects like neutron stars, quark stars, 
provide natural sites where many of the theoretical
conjectures about the various phases of quark matter can be tested. 
The latter, in the present context, is more relevant here, as we
study the possibility of para-ferro phase transition in dense quark
system interacting via one gluon exchange. 

The original idea about para-ferro phase transition in quark matter
was proposed recently in \cite{tatsumi00} where the possibility of
Bloch like phase transition \cite{bloch29} was studied and it 
was shown that spin polarized quark matter might exist at low density 
\cite{niegawa05}. The underlying mechanism of such a phase transition 
is analogous to what was originally proposed for the degenerate electron 
gas \cite{bloch29}. There, for Coulomb interaction, it was shown that 
the exchange correction to the energy is attractive which at the low density
wins over the kinetic energy giving rise to a ferromagnetic state \cite{bloch29}. 
In \cite{tatsumi00}, a variational calculation
has been performed to show that it is indeed possible to have a spin
polarized quark matter at low density of strange quark system, while
for light quark it never happens \cite{tatsumi00}. Similar difference
of the light and strange quark matter,
albeit in a different context, was observed earlier \cite{chin79prl}. 
However, in \cite{niegawa05}, it was shown that both the light and heavy flavor 
systems can exhibit such phase transitions although the critical density
for the strange matter is higher than the light quark systems.
Such investigations, have also been performed in
\cite{nakano03,ohnishi07,inui07,tatsumi08} and also in 
\cite{tatsumi09,tat_09} where the calculation has been
extended to include thermal effects. The Bloch like phase transition, 
for strange quark, has also been reconfirmed in \cite{pal09}.

One shortcoming of all these works including \cite{pal09}, has been that
the calculations were restricted to the Hartree Fock
level and the terms beyond the exchange diagrams, commonly termed
as correlation energy \cite{freed77,chin77,kap_book,wal_book,pal09_gse}
were ignored. Without such corrections, however, the calculations
are known to remain 
incomplete as the higher order terms are plagued with infrared
divergences arising out of the exchange of massless gluons, 
indicating the failure of the naive perturbation series. 
We know that this problem can be cured by reorganizing the perturbation
theory where a particular class of diagrams, {\em viz.} the
bubbles are resummed in order to obtain a finite result. Originally,
as is well known, this was done by Gell-Mann and Brueckner \cite{gellmann57}
while calculating the ground state energy of degenerate electron gas. 
The contribution of the bubbles involve terms of ${\cal O}(g^4\ln g^2)$ indicating 
non-perturbative nature of the correction 
\cite{sawada57,pines58,pines_book,perdew92}.

In the present work, as announced, we calculate the spin susceptibility $(\chi)$
of dense quark system with corrections due to correlations {\em i.e.}
containing terms upto ${\cal O}(g^4\ln g^2)$. This
requires the knowledge of the ground state energy (GSE) of spin polarized
matter with inclusion of bubble diagrams. The GSE of the polarized quark matter
has been calculated only recently in \cite{pal09_gse}
which is the starting point of the present paper. This work is
very similar to that of Brueckner and Swada \cite{bruek58}
and those of \cite{shastry77,shastry78},
applied to the case of QCD matter. Unlike, degenerate electron gas, however, we have both the electric and magnetic interactions and the calculation is performed relativistically, while the non-relativistic results appear as a limit. 

The spin susceptibility $\chi$, for quark matter upto ${\cal O}(g^2)$
has already been calculated in Ref.\cite{tatsumi00} which we only briefly
discuss. Subsequently, the non-fermi liquid corrections to $\chi$ has also 
been studied in \cite{tatsumi09,tat_09}. These studies provide further motivation
to undertake the present endeavor to include correlation corrections, without
which, as mentioned already, the perturbative evaluation of ${\chi}$ remains incomplete. 
In addition, we also calculate incompressibility and sound velocity for spin
polarized quark matter with corrections due to correlations which involve 
evaluation of single particle energy at the Fermi surface. These quantities 
are of special interests for applications to astrophysics. Moreover, 
we also evaluate the exchange energy density at non-zero temperature 
and determine the corresponding corrections to the spin susceptibility.

The plan of the paper is as follows.
In Sec. II we calculate spin susceptibility with correlation 
correction for degenerate quark matter. Analytic expressions are 
presented both in ultra-relativistic (UR) and non-relativistic (NR) limit.
In Sec. III, we evaluate exchange energy density and spin-susceptibility 
at non-zero temperature.
In Sec. IV we summarize and conclude. Detailed 
expressions of the intermediate expressions, from which $\chi$ is
derived, have been relegated to the Appendix.

\section{Spin susceptibility}

The spin susceptibility of quark matter is determined by the change in 
energy of the system as its spins are polarized \cite{bruek58}. We introduce a 
polarization parameter ${\xi}= (n^+_q-n^-_q)/n_q$ with the condition 
$ 0\le{\xi}\le 1$, where $n_q^+$ and $n_q^-$ correspond to densities 
of spin-up and spin-down quarks respectively, and 
$n_{q}=n_{q}^{+}+n_{q}^{-}$ denotes total quark density. 
The Fermi momenta in the spin-polarized quark matter then 
are $p_{f}^{+}=p_{f}(1+{\xi})^{1/3}$ and 
$p_{f}^{-}=p_{f}(1-{\xi})^{1/3}$, where $p_{f}=(\pi^2n_{q})^{1/3}$, 
is the Fermi momentum of the unpolarized matter $({\xi}=0)$.
In the small $\xi$ limit, the ground state energy behaves like
\cite{tatsumi00}

\beq\label{xi_expan}
E(\xi)&=& E(\xi=0)+\frac{1}{2}\beta_s\xi^2+{\cal O}(\xi^4).
\eeq
Here, $\beta_s=\frac{\del^2 E}{\del\xi^2}{\Big|}_{\xi=0}$, defined to be 
the spin stiffness constant in analogy with \cite{pal09_gse,perdew92}.
The spin susceptibility $\chi$ is proportional to the inverse of 
the spin stiffness, mathematically $\chi=2 \beta_s^{-1}$ \cite{perez09}.
It is to be noted that in Eq.(\ref{xi_expan}), the first term corresponds to
unpolarized matter energy.

Now, the leading contributions to the 
ground state energy are given by
the three terms {\em {viz.}} kinetic, exchange and 
correlation energy density \cite{pal09_gse} {\rm i.e.}
 
\beq
E &=& E_{kin}+E_{ex}+E_{corr}.
\eeq

The total kinetic energy density for spin-up 
and spin-down quark becomes \cite{tatsumi00,pal09}

\beq\label{rel_kin}
E_{kin}&=&\frac{3}{16\p^2}
\left\{p_f(1+\xi)^{1/3}\sqrt{p_f^2(1+\xi)^{2/3}+m_q^2}
\left[2p_f^2(1+\xi)^{2/3}+m_q^2\right]\right.\nn\\&&\left.
-m_q^4\ln\left(\frac{p_f(1+\xi)^{1/3}+\sqrt{p_f^2(1+\xi)^{2/3}+m_q^2}}
{m_q}\right)
+[\xi\rightarrow -\xi]\right\},
\eeq
where $m_q$ is the quark mass.

The exchange energy density $E_{ex}$ have been calculated 
in ref.\cite{pal09} within Fermi liquid theory approach. One can also directly 
evaluate the two loop diagram \cite{tatsumi00} to obtain 

\beq
E_{ex}^{nf}&=&\frac{9}{2}\sum_{s=\pm}\int\int\frac{d^3p}{(2\p)^3}
\frac{d^3p'}{(2\p)^3}\theta(p_f^s-|p|)\theta(p_f^s-|p'|)f_{pp'}^{nf},
\label{ex_nf}\\
E_{ex}^{f}&=& 9 \int\int\frac{d^3p}{(2\p)^3}
\frac{d^3p'}{(2\p)^3}\theta(p_f^{+}-|p|)\theta(p_f^{-}-|p'|)f_{pp'}^{f},
\label{ex_f}
\eeq 

where $f_{pp'}^{nf}$ and $f_{pp'}^{f}$ stands for non-flip $(s=s')$ and 
flip $(s=-s')$ forward scattering amplitude given in 
\cite{tatsumi00,pal09,pal09_gse}. Here, $E_{ex}=E_{ex}^{nf}+E_{ex}^{f}$ 
can be estimated numerically. However, analytical evaluation of these
integrals is possible in the ultra-relativistic and non-relativistic limits  
as reported in \cite{tatsumi00,pal09,pal09_gse}.

The next higher order correction to the ground state energy beyond the
exchange term is the correlation energy $E_{corr}$ 
\cite{freed77,chin77,kap_book,wal_book}. The detailed calculation 
of correlation energy for spin polarized matter have been derived in \cite{pal09_gse}
which we quote here:

\beq\label{cor_eng2}
E_{corr} & \simeq & \frac{1}{(2\p)^3}\frac{1}{2}
\int_0^{\p/2}\sin^2\theta_E {\rm d}\theta_E
\left\{\Pi_L^2\left[\ln\left(\frac{\Pi_L}{\veps_f^2}\right)-\frac{1}{2}\right]
+2\Pi_T^2\left[\ln\left(\frac{\Pi_T}{\veps_f^2}\right)-\frac{1}{2}\right]
\right\},
\eeq

with $\theta_E=\tan^{-1}(|k|/k_0)$. The relevant
$\Pi_L$ and $\Pi_T$ are determined to be \cite{pal09_gse}

\beq
\Pi_L &=& \frac{g^2}{4\p^2}\sum_{s=\pm}\frac{p_f^s\veps_f^s}{\sin^2\theta_E}
\left[1-\frac{\cot\theta_E}{v_f^s}
\tan^{-1}\left(v_f^s\tan\theta_E \right)\right],
\label{piL_polar}\\
\Pi_T &=& \frac{g^2}{8\p^2}\sum_{s=\pm}{p_f^s}^2\cot\theta_E
\left[-\frac{\cot\theta_E}{v_f^s}+
\left(1+\frac{\cot^2\theta_E}{{v_f^s}^2}\right)
\tan^{-1}\left(v_f^s\tan\theta_E \right)\right].
\label{piT_polar}
\eeq

The spin susceptibility is given by\cite{tatsumi00} 

\beq
\chi^{-1}&=&\frac{1}{2}\frac{\del^2 E(\xi)}{\del \xi^2}{\Big|}_{\xi=0}.
\eeq

We have $\chi^{-1} \equiv \chi^{-1}_{kin} +\chi^{-1}_{ex} +\chi^{-1}_{corr}$.
The kinetic and exchange contribution have been evaluated in 
ref.\cite{tatsumi00}, is given by 

\beq
\chi_{kin}^{-1}&=&\frac{p_f^5}{6\pi^2\veps_f}.
\label{chi_kin}\\
\chi_{ex}^{-1}&=&-\frac{g^2p_f^4}{18\pi^4}
\left\{2-\frac{6p_f^2}{\veps_f^2}
-\frac{3p_f}{\veps_f^3}\left[p_f\veps_f-m_q^2
\ln\left(\frac{p_f+\veps_f}{m_q}\right)\right]
+\frac{2p_f^2}{\veps_f^2}\left[1+\frac{2m_q}{3(p_f+m_q)}\right]\right\}.
\label{chi_ex}\nn\\
\eeq 

To determine the correlation correction to spin susceptibility,
we expand curly braces terms of Eq.(\ref{cor_eng2}) in powers of the 
polarization parameter $\xi$, which gives

\beq\label{cor_expan}
\Pi_L^2\left[\ln\left(\frac{\Pi_L}{\veps_f^2}\right)-\frac{1}{2}\right]
+2\Pi_T^2\left[\ln\left(\frac{\Pi_T}{\veps_f^2}\right)-\frac{1}{2}\right]
&=&({\cal A}_{0L}+{\cal B}_{0T})+\xi^2({\cal A}_{1L}+{\cal B}_{1T})
+{\cal O}(\xi^4).\nn\\
\eeq
Here, ${\cal A}_{0L}$ and ${\cal B}_{0T}$ correspond to unpolarized matter 
term and the detailed expressions of ${\cal A}_{1L}$ and ${\cal B}_{1T}$ 
are given in the Appendix. $\chi_{corr}^{-1}$ is 

\beq\label{chi_corr}
\chi_{corr}^{-1} &=& \frac{1}{2}\frac{\del^2 E_{corr}(\xi)}
{\del \xi^2}{\Big|}_{\xi=0}\nn\\
&\simeq& \frac{1}{(2\p)^3}\frac{1}{2}
\int_0^{\p/2}\sin^2\theta_E {\rm d}\theta_E ({\cal A}_{1L}+{\cal B}_{1T}).
\eeq

From the above expression and with the help of the expression presented in
the Appendix, $\chi_{corr}^{-1}$ can be estimated numerically. Results
for the two limiting cases however can be obtained analytically as
we present in the following two sub-sections.

\subsection {Ultra-relativistic limit}

In the ultra-relativistic limit, the kinetic, exchange and correlation 
energies are \cite{pal09_gse}

\beq
E_{kin}^{ur}&=&\frac{3p_{f}^4}{8\pi^2}
\left[(1+{\xi})^{4/3}+(1-{\xi})^{4/3}\right],\nn\\
E_{ex}^{ur}&=&\frac{g^2}{32\pi^4}p_{f}^4\left[(1+{\xi})^{4/3}+
(1-{\xi})^{4/3}+2(1-{\xi}^2)^{2/3}\right],\nn\\
E_{corr}^{ur} &=& \frac{g^4\ln g^2}{2048 \p^6}p_f^4
[(1+\xi)^{4/3}+(1-\xi)^{4/3}+2(1-\xi^2)^{2/3}].
\label{ur_eng}
\eeq

With the help of Eq.(\ref{xi_expan}), each energy contribution to the 
susceptibility is

\beq\label{ur_contri}
\chi_{kin}^{-1} &=& \frac{p_f^4}{6\p^2}\nn\\
\chi_{ex}^{-1} &=& -\frac{g^2 p_f^4}{36\p^4}\nn\\
\chi_{corr}^{-1} &=& -\frac{g^4 p_f^4}{2304\p^6}(\ln r_s-0.286).
\eeq

with $r_s=g^2(\frac{3\pi}{4})^{1/3}$. From Eq.(\ref{ur_contri}), sum of 
all the contribution to the susceptibility can be written as \cite{pal09_gse}

\beq\label{ur_suscep}
\chi^{ur} &=& \chi_P[1-\frac{g^2}{6\p^2}-\frac{g^4}{384\p^4}(\ln r_s-0.286)]^{-1},
\eeq

where $\chi_P$ is the non-interacting susceptibility 
\cite{shastry77,shastry78}.

\subsection {Non-relativistic limit}

Now, we go to the non-relativistic limit to calculate spin-susceptibility
in order to compare our results with those of dense
electron gas \cite{bruek58,shastry77,shastry78,herri_book} interacting via. 
static Coulomb potential. In this limit, kinetic and exchange 
energy densities are \cite{tatsumi00,pal09,pal09_gse}:

\beq
E_{kin}^{nr}&=&\frac{3p_{f}^5}{20\pi^2m_{q}}
\left[(1+\xi)^{5/3}+(1-\xi)^{5/3}\right],
\label{kin_nr}\nn\\
E_{ex}^{nr}&=&-\frac{g^2}{8\pi^4}p_{f}^4
\left[(1+\xi)^{4/3}+(1-\xi)^{4/3}\right].
\label{ex_nr}
\eeq

The contribution to the susceptibility from kinetic and exchange 
energy density yields

\beq
\chi_{kin}^{-1}&=&\frac{p_f^5}{6\pi^2m_q}\nn\\
\chi_{ex}^{-1}&=&-\frac{g^2p_f^4}{18\pi^4}.
\eeq

We want to calculate the contribution to the spin-susceptibility 
beyond the exchange correction. For this we first evaluate the 
correlation energy in this limit.

The dominant contribution to the correlation energy 
is found to be,

\beq\label{nr_corr}
E_{corr}^{nr}&=&-\frac{{\lambda^2} {p_f^5}}{\pi^4 m_q}
\int_{\lambda^{1/2}}^{k_c}\frac{{\rm d}k'}{k'}
\int_0^{\infty}x{\rm d}x\sum_{s=\pm}f(s)
\Big[1-\frac{x^s}{2}\ln\left(\frac{x^s+1}{x^s-1}\right)\Big]\sum_{s'=\pm}
\theta(1-x^{s'}),
\eeq

where $\lambda=(g^2 m_q)/(8\pi p_f)$, $f(s=\pm)=(1\pm \xi)^{1/3}$, $x=x^s f(s)$, 
$x^s=(k_0 m_q)/(p_f^s k)$ and $k'=k/p_f$. For $s=s'$ one obtains: 

\beq\label{corr_nf}
E_{corr}^{nr, s=s'}
&\simeq& \frac{g^4\ln g^2}{(2\pi)^6}\frac{1}{3}m_q p_f^3(1-\ln 2),
\eeq

Note that, here, the correlation energy is independent of 
spin-polarization $\xi$. It is seen, that for the spin parallel 
interactions $\xi$-dependent terms 
contribute with opposite sign and cancels each other.
For $s=-s'$, the integral on $x$ takes the form

\beq\label{xi2_term1}
I&=&\int_0^{\infty}x{\rm d}x\left\{(1+\frac{1}{3}\xi)
\left[1-\frac{1}{2}x(1-\frac{1}{3}\xi)
\ln{\Big|}\frac{x(1-\frac{1}{3}\xi)+1}
{x(1-\frac{1}{3}\xi)-1}{\Big|}\right]
\theta[1-x(1+\frac{1}{3}\xi)]+(\xi \ra -\xi)\right\}.\nn\\
\eeq

Expanding $\ln$ in terms of $\xi$ and retain upto ${\cal O}(\xi^2)$ 
we have

\beq\label{xi2_term2}
I&\simeq& \frac{2}{3}\left[(1-\ln 2)-\frac{1}{6}\xi^2\right].
\eeq

Using Eq.(\ref{nr_corr}), (\ref{xi2_term1}) and (\ref{xi2_term2}) we have

\beq\label{corr_f}
E_{corr}^{nr,s=-s'}&\simeq& \frac{g^4\ln g^2}{128\pi^6}
\frac{1}{3}m_q p_f^3\left[(1-\ln 2)-\frac{1}{6}\xi^2\right].
\eeq

It is to be mentioned that similar expressions for degenerate 
electron gas interacting via. static Coulomb potential can be found in 
ref.\cite{herri_book}. From Eq.(\ref{corr_nf}) and (\ref{corr_f}), 
it is clear that spin anti-parallel states are attractive in contrast
to the parallel states due to Pauli exclusion principle.
In this limit the correlation contribution to the susceptibility is
found to be,

\beq
\chi_{corr}^{-1}&=&-\frac{g^4\ln g^2}{2304\pi^6}m_q p_f^3.
\eeq

The total susceptibility is given by

\beq
\chi^{nr}&=&\chi_{P}\left[1-\frac{g^2}{3\pi^2}\frac{m_q}{p_f}
-\frac{g^4\ln g^2}{384\pi^4}\frac{m_q^2}{p_f^2}\right]^{-1}.
\eeq

\vskip 0.2in
\begin{figure}[htb]
\begin{center}
\resizebox{9.0cm}{7.2cm}{\includegraphics[]{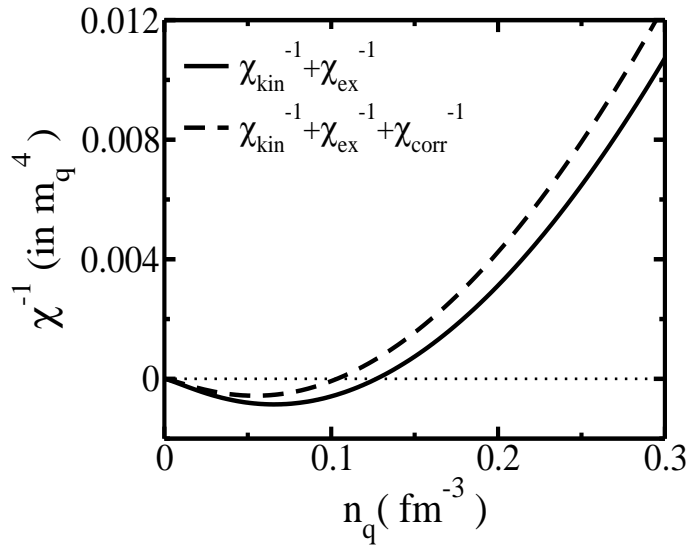}}
\caption{Density dependence of inverse spin susceptibility.}
\label{chi_comp}
\end{center}
\end{figure}


In Fig.(\ref{chi_comp}) we plot inverse spin susceptibility 
which is valid for all the kinematic regimes.
It shows $\chi^{-1}$ changes its sign at the density $\sim 0.12 {\rm fm^{-3}}$
without correlation correction and when we include the correlation effect
its sign changes at $\sim 0.1 {\rm fm^{-3}}$. This is equivalent to what
happens to the ground state energy as a function of $\xi$. It is needless
to mention that this change of sign correspond to the para-ferro phase 
transition in dense quark system. The parameter set used here are same as
those of \cite{tatsumi00,chin79prl,pal09,pal09_gse}.

\subsection {Incompressibility and Sound velocity}

Once we have the expressions for the total energy density, the 
incompressibility $(K)$ and sound velocity $(c_1)$ can be determined. 
The incompressibility $K$ is defined by the second 
derivative of the total energy density with respect to the number 
density $n_q$, which is given by\cite{pal09}

\beq
K&=&9n_{q}\frac{\del^2 E}{\del n_{q}^2}.
\eeq

Since, there are two Fermi surfaces corresponding to spin-up $(+)$ and 
spin-down $(-)$ states, such that $E\equiv E(n_{q}^{+},n_{q}^{-})$. 
We have \cite{pal09}

\beq\label{delE_deln}
\frac{\del E}{\del n_{q}}&=&\frac{\del E}{\del n_{q}^{+}}
\frac{\del n_{q}^{+}}{\del n_{q}}+\frac{\del E}{\del n_{q}^{-}}
\frac{\del n_{q}^{-}}{\del n_{q}}\nn\\
&=&\frac{1}{2}\left[(1+{\xi})\mu^{+}+(1-{\xi})\mu^{-}\right].
\eeq

The single particle energy at the Fermi surface or chemical potential of 
spin-up quark turns out to be

\beq\label{up_chpot}
\mu^{+,ur}&=&\mu_{kin}^{+}+\mu_{ex}^{+}+\mu_{corr}^{+}\nn\\
&=&p_f^++\frac{g^2}{12\pi^2}\left(p_f^{+}+\frac{p_f^{+2}}{p_f^-}\right)
+\frac{g^4\ln g^2}{768\pi^4}\left(p_f^{+}+\frac{p_f^{+2}}{p_f^-}\right).
\eeq

Similarly, $\mu^{-,ur}$ can be obtained by replacing $p_f^{\pm}$ with 
$p_f^{\mp}$ in Eq.(\ref{up_chpot}). In ref.\cite{pal09}, 
chemical potential is determined within Fermi liquid theory 
approach upto ${\cal O}(g^2)$. We, however, here calculate $\mu^{\pm}$ 
with different approach upto ${\cal O}(g^4 \ln g^2)$.

Using Eq.(\ref{delE_deln}) and Eq.(\ref{up_chpot}), the incompressibility
becomes

\beq\label{incomp}
K^{ur}&=&\frac{3}{2}p_f\left\{[(1+{\xi})^{4/3}+(1-{\xi})^{4/3}]
+\frac{g^2}{12\pi^2}[(1+{\xi})^{4/3}+(1-{\xi})^{4/3}
+2(1-{\xi}^2)^{2/3}]\right.\nn\\&&\left.
+\frac{g^4\ln g^2}{768\pi^4}[(1+{\xi})^{4/3}+(1-{\xi})^{4/3}
+2(1-{\xi}^2)^{2/3}]\right\}.
\eeq

Another interesting quantity would be to calculate the first sound 
velocity  which is given by the first derivative of pressure with 
respect to energy density. Mathematically \cite{pal09},

\beq
c_1^2&=&\left[\frac{(1+{\xi})n_{q}^{+}\frac{\del \mu^{+}}{\del n_{q}^{+}}
+(1-{\xi})n_{q}^{-}\frac{\del \mu^{-}}{\del n_{q}^{-}}}
{(1+{\xi})\mu^{+}+(1-{\xi})\mu^{-}}\right].
\eeq

From Eq.(\ref{up_chpot}), we have

\beq
\frac{\del \mu^+}{\del n_q^+}&=&\frac{2\pi^2}{3p_f^2(1+\xi)^{2/3}}
\left\{1+\frac{g^2}{12\pi^2}
\left[\frac{(1+{\xi})^{2/3}-(1-{\xi})^{2/3}}{(1+{\xi})^{2/3}}\right]
\right.\nn\\&&\left.
+\frac{g^4\ln g^2}{768\pi^4}
\left[\frac{(1+{\xi})^{2/3}-(1-{\xi})^{2/3}}{(1+{\xi})^{2/3}}\right]\right\}.
\eeq

The second and last term in the curly braces corresponds to 
exchange and correlation contribution respectively. 
Similarly, $\del \mu^-/\del n_q^-$ can be 
obtained by replacing $\xi$ with $-\xi$. Using $n_q^{\pm}$, 
$\mu^{\pm}$ and $\del \mu^{\pm}/\del n_q^{\pm}$, 
we calculate the sound velocity in terms of $\xi$.
Numerically, for unpolarized matter $c_1=0.46$, 
while for complete polarized matter $c_1=0.54$, 
which is below the causal value $1/\sqrt{3}=0.57$ at the high density limit.

\vskip 0.2in
\begin{figure}[htb]
\begin{center}
\resizebox{9.0cm}{7.0cm}{\includegraphics[]{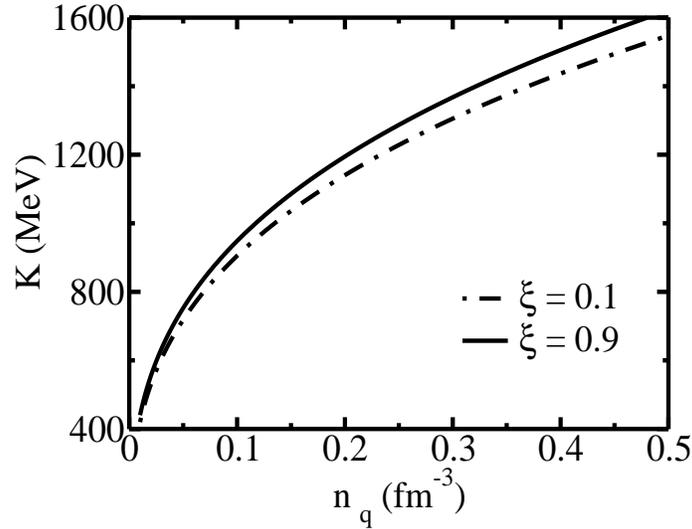}}
\caption{Incompressibility $K$ in the spin polarized quark matter.}
\label{compres}
\end{center}
\end{figure}


In Fig(\ref{compres}), we plot the density dependencies of the 
incompressibility with correlation correction. This shows for higher 
value of the order parameter $\xi$, the incompressibility 
becomes higher for the same value of density. Thus numerical 
values of incompressibility and sound velocity shows that
equation of state for polarized quark matter is
stiffer than the unpolarized one \cite{pal09}.

\section{Susceptibility at non-zero Temperature}

In this section we calculate the exchange energy density $E_{ex}$ at 
low-temperature $(T<<\veps_f)$, for which we replace  
$\theta(p_f^{\pm}-|p|)$ of Eqs.(\ref{ex_nf}-\ref{ex_f}) with proper Fermi distribution function. In the ultra-relativistic limit, the angular averaged interaction parameter is given by \cite{pal09}

\beq\label{int_ur}
f_{pp'}^{ur}&=&\frac{g^2}{9pp'}
\int \frac{\rm d\Omega_{1}}{4\pi}\int \frac{\rm d\Omega_{2}}{4\pi}
\Big[1+({\hat p}\cdot {\hat s})({\hat p'}\cdot {\hat s'})\Big]
\eeq

The spin non-flip contribution to the exchange energy density is

\beq\label{exnf_ur}
E_{ex}^{nf}&=&\frac{9}{2}\sum_{s=\pm}\int\int\frac{d^3p}{(2\p)^3}
\frac{d^3p'}{(2\p)^3}f_{pp'}^{nf}~n_{p}^s(T)~n_{p'}^s(T)\nn\\
&\simeq&\frac{g^2}{32\pi^4}p_{f}^4\Big[(1+{\xi})^{4/3}+
(1-{\xi})^{4/3}\Big]+\frac{g^2}{48\pi^2}T^2p_f^2
\Big[(1+{\xi})^{2/3}+(1-{\xi})^{2/3}\Big].
\eeq
Here $n_{p(p')}^{s}(T)$ is the Fermi distribution function.

Similarly, $E_{ex}^{f}$ can be evaluated. The total $E_{ex}^{ur}$
at low temperature is found to be

\beq\label{ex_urT}
E_{ex}^{ur}&\simeq&\frac{g^2}{32\pi^4}p_{f}^4\Big[(1+{\xi})^{4/3}+
(1-{\xi})^{4/3}+2(1-{\xi}^2)^{2/3}\Big]\nn\\
&+&\frac{g^2}{24\pi^2}T^2p_f^2
\Big[(1+{\xi})^{2/3}+(1-{\xi})^{2/3}\Big].
\eeq

The kinetic energy density can be written as

\beq\label{kin_urT}
E_{kin}^{ur}&\simeq&\frac{3p_{f}^4}{8\pi^2}
\Big[(1+{\xi})^{4/3}+(1-{\xi})^{4/3}\Big]+\frac{3T^2p_f^2}{4}
\left[(1+{\xi})^{2/3}+(1-{\xi})^{2/3}\right].
\eeq

From Eq.(\ref{xi_expan}) each energy contribution to the susceptibility is

\beq
\chi_{kin}^{-1}&=&\frac{p_f^4}{6\pi^2}\Big(1-\frac{\pi^2 T^2}{p_f^2}\Big)\nn\\
\chi_{ex}^{-1}&=&-\frac{g^2 p_f^4}{36\pi^4}\Big(1+\frac{\pi^2 T^2}{3p_f^2}\Big).
\label{exsus_T}
\eeq

It is to be noted that the $T$ independent terms of the above expressions 
are identical with those Eqs.(\ref{ur_eng}-\ref{ur_contri}). 
Thus the susceptibility at non-zero temperature is given by

\beq\label{sus_urT}
\chi^{ur}&=&\chi_{P}\Big[1-\frac{g^2}{6\pi^2}
\Big(1+\frac{4\pi^2T^2}{3p_f^2}\Big)\Big]^{-1}.
\eeq

In the non-relativistic limit the interaction parameter takes the 
following form \cite{tatsumi00,pal09}

\beq
f_{pp'}^{nr}=-\frac{2g^2}{9}\Big[\frac{1+s\cdot s'}{|p-p'|^2}\Big].
\eeq

For spin anti-parallel interaction $s=-s'$, then $f_{pp'}^{nr}=0$. 
Thus the contribution due to the scattering of quarks with unlike 
spin states vanishes and  the dominant contribution to energy density 
comes from the parallel spin states $(s=s')$. Performing the angular 
integration of Eq.(\ref{ex_nf}), the exchange energy density 
upto term ${\cal O}(T^2)$ becomes 

\beq
E_{ex}^{nr}&=&-\frac{g^2}{4\pi^4}\sum_{s=\pm}\int p~d{\rm p}~n_p^s(T)
\int p'~d{\rm p'}~n_{p'}^s(T)\ln\Big|\frac{p+p'}{p-p'}\Big|\nn\\
&\simeq&-\frac{g^2}{8\pi^4}p_{f}^4\Big[(1+\xi)^{4/3}+(1-\xi)^{4/3}\Big]
-\frac{g^2}{8\pi^2}T^2 m_{q}p_{f}\Big[(1+\xi)^{1/3}+(1-\xi)^{1/3}\Big].
\eeq

The kinetic energy density is found to be

\beq
E_{kin}^{nr}&\simeq&\frac{3p_{f}^5}{20\pi^2m_{q}}
\Big[(1+\xi)^{5/3}+(1-\xi)^{5/3}\Big]
+\frac{T^2p_f^2}{2}\Big[(1+\xi)^{2/3}+(1-\xi)^{2/3}\Big].
\eeq

Separate contribution from kinetic and exchange energy to susceptibility 
becomes

\beq
\chi_{kin}^{-1}&=&\frac{p_f^5}{6\pi^2m_{q}}
\Big(1-\frac{2\pi^2m_{q}T^2}{3p_f^3}\Big)\nn\\
\chi_{ex}^{-1}&=&-\frac{g^2p_f^4}{18\pi^4}
\Big(1-\frac{\pi^2m_{q}T^2}{2p_f^3}\Big).
\eeq

Thus, at low temperature the susceptibility turns out to be

\beq
\chi^{nr}&=&\chi_{P}\Big[1-\frac{g^2m_{q}}{3\pi^2p_f}
\Big(1+\frac{\pi^2m_{q}T^2}{6p_f^3}\Big)\Big]^{-1}.
\eeq

\section{Summary and conclusion}

In this work we have derived the spin susceptibility
for degenerate quark matter with corrections due to correlation contributions. 
Analytic expressions for susceptibility are also derived both in 
the ultra-relativistic and the non-relativistic limit. It is observed that
at low density susceptibility changes sign and becomes negative suggesting
the possibility of ferromagnetic phase transition.  In addition, we also derive
single particle energy, sound velocity and incompressibility
upto ${\cal O}(g^4 \ln g^2)$. As far as the equation of state 
is concerned, in the present model, we find that equation of state 
for polarized matter is stiffer than that of unpolarized one. We also 
determine the exchange energy and susceptibility at non-zero temperature
of the spin polarized quark matter.

\newpage
\section {Appendix}

To calculate the correlation contribution to the spin susceptibility
we have from Eq.(\ref{cor_expan})

\beq
{\cal A}_{1L} &=& -\frac{g^4p_f^2\sec^4\theta_E\csc^6\theta_E}
{1152\pi^4\veps_f^3(m_q^2+p_f^2\sec^2\theta_E)^2}\nn\\&\times&
\ln\left\{\frac{g^2\csc^2\theta_E}{2\pi^2\veps_f}
\Big[p_f-\veps_f\cot\theta_E\tan^{-1}(v_f\tan\theta_E)
\Big]\right\}\nn\\&\times&
\Big\{64\veps_f^5\cos^2\theta_E\tan^{-1}(v_f\tan\theta_E)
(p_f^2+m_q^2\cos^2\theta_E)^2\nn\\&-&
2p_f\veps_f^2\sin 2\theta_E\tan^{-1}(v_f\tan\theta_E)
\{12m_q^6+51m_q^4p_f^2\nn\\&+&
m_q^4(4m_q^2+5p_f^2)\cos 4\theta_E
+68m_q^2p_f^4+4m_q^2(4m_q^4+10m_q^2p_f^2+7p_f^4)\cos 2\theta_E
+32p_f^6\}\nn\\&+&
4p_f^2\veps_f\sin^2\theta_E[6m_q^6+29m_q^4p_f^2
+m_q^4(2m_q^2+3p_f^2)\cos 4\theta_E+36m_q^2p_f^4\nn\\&+&
4m_q^2(2m_q^4+4m_q^2p_f^2+3p_f^4)\cos 2\theta_E+16p_f^6]\Big\}
\label{A_1L}\\
{\cal B}_{1T} &=&\frac{g^4p_f^2\cot^2\theta_E\csc^4\theta_E}
{1152\pi^4\veps_f^3(m_q^2\cos^2\theta_E+p_f^2)}\nn\\&\times&
\ln\left\{\frac{g^2\cot\theta_E\csc^2\theta_E}{8\pi^2\veps_f^2}
\Big[2\tan^{-1}(v_f\tan\theta_E)(m_q^2\cos^2\theta_E+p_f^2)
-p_f\veps_f\sin 2\theta_E\Big]\right\}\nn\\&\times&
\Big\{-32\veps_f^3\tan^{-1}(v_f\tan\theta_E)
(m_q^2\cos^2\theta_E+p_f^2)^2\nn\\&-&
8p_f^2\veps_f[m_q^4+p_f^4+m_q^2p_f^2(1+\cos^2\theta_E)]
\sin^2 2\theta_E+2p_f\tan^{-1}(v_f\tan\theta_E)\sin 2\theta_E
\nn\\&\times&
[8m_q^6+31m_q^4p_f^2+m_q^4p_f^2\cos 4\theta_E+36m_q^2p_f^4
+4m_q^2(2m_q^4+4m_q^2p_f^2+3p_f^4)\cos 2\theta_E+16p_f^6]\Big\}.\nn\\
\label{B_1T}
\eeq

with $v_f=p_f/\veps_f$.

\vskip 0.2in
{\bf Acknowledgments}\\

The authors would like to thank S.Mallik for the critical reading of 
the manuscript and S.Biswas for his help with the 
numerical calculations. The author K.Pal would like to special 
thanks T.Tatsumi for his valuable suggestions.

\end{document}